\documentclass[twocolumn,aps,prb,reprint,superscriptaddress,longbibliography]{revtex4-2}

\usepackage{braket}
\usepackage{bm}
\usepackage{multirow}
\usepackage{amsmath,amsfonts,amssymb}
\usepackage[english]{babel}
\usepackage{graphicx}
\usepackage{hyperref}
\usepackage{dsfont} 
\hypersetup{
     colorlinks   = true,
     citecolor    = blue
}

\begin{document}

\title{Designing Flat Bands and Pseudo-Landau Levels in GaAs with Patterned Gates}

\author{Pierre A. Pantale\'{o}n}
\email{pierre.pantaleon@imdea.org}
\affiliation{Imdea Nanoscience, Faraday 9, 28015 Madrid, Spain}
\affiliation{Department of Physics, Colorado State University, Fort Collins, Colorado 80523, USA}
\author{Zhen Zhan}
\affiliation{Imdea Nanoscience, Faraday 9, 28015 Madrid, Spain}
\author{Siddhartha E. Morales}
\affiliation{
 Depto. de Sistemas Complejos, Instituto de F\'{i}sica, Universidad Nacional Aut\'{o}noma de M\'{e}xico (UNAM). Apdo. Postal 20-364, 01000 M\'{e}xico D.F., M\'{e}xico
 }
 \author{Gerardo G. Naumis}
\email{naumis@fisica.unam.mx}
\affiliation{
 Depto. de Sistemas Complejos, Instituto de F\'{i}sica, Universidad Nacional Aut\'{o}noma de M\'{e}xico (UNAM). Apdo. Postal 20-364, 01000 M\'{e}xico D.F., M\'{e}xico
 }%

\begin{abstract}
We investigate the electronic properties of two-dimensional electron gases (2DEGs) subjected to a periodic patterned gate. By incorporating the superlattice (SL) potential induced by patterning into the Schr\"{o}dinger equation, we develop a methodology for obtaining exact analytical solutions.\ These solutions enable us to construct a comprehensive phase diagram illustrating the emergence of narrow bands and pseudo-Landau levels driven by the SL potential.\ To complement the analytical approach, we employ a standard plane-wave formalism to track the evolution of the band structure as the SL strength increases.\ By breaking the inversion symmetry of the SL potential, we found a nontrivial Berry curvature. Furthermore, we introduce a self-consistent Hartree screening to account for the interplay between the SL potential and electronic interactions.\ Our findings not only reveal the emergence of a non-trivial quantum geometry and a competition between SL strength and electron-electron interactions, but also highlight the value of exact analytical solutions for understanding and engineering electronic phases in patterned 2DEG systems.

\end{abstract}
\maketitle

\section{Introduction}

Many years ago, Albrecht et.\ al.\ investigated high mobility two-dimensional electron systems
subjected to a periodic superlattice (SL) potential using GaAs hetero-junctions~\cite{VonKlitzing1999}. In these systems, the electronic properties were found to be determined by the successive quantum wells rather than by the individual semiconductor layers~\cite{Duffield_1986}. More recently, the interest on such kind of systems has been revived in the context of emulating 2D materials and 2D moir\'e modulated materials in semiconductors~\cite{wang2024Tuning,wang2024Lateral,Gomes2012Designer} and graphene heterostructures~\cite{tan2024designing,zhan2024designing,ghorashi2023multilayer,ghorashi2023topological,krix2023patterned}. One of the primary motivations behind this revival is the potential to produce highly correlated many-body phases~\cite{Chubukov2017Super,Phillips1998,Raghu2011Super}, similar to those observed in magic-angle twisted graphene~\cite{Cao2018,Jarrillo2021_Fractional,park2021}. As is well known, flat bands play a key role in enabling this behavior~\cite{Tian2023Evidence,Chen2024Ginzburg,Cano2021Exact}. These flat bands are associated with pseudo-Landau levels~\cite{Cano2021Exact,popovF2021,2021Yarden} and can be considered as a form of topological solitons~\cite{Elias_2023}.

Here we propose a strategy to achieve a non trivial quantum geometry of flat bands in semiconductor heterostructures, opening the possibility of doing superlattice engineering as in graphene systems~\cite{forsythe2018band,huber2020gate,barcons2022engineering,barcons2022engineering,Wang2024Dispersion}.\ We remark that flat bands in semiconductor
 heterostructures were previously found in some other works~\cite{ Du2021Observation,Nadvorník_2012,Sushkov2013Topological}.\ However, our aim here is to go beyond such results to show how the interplay between the lattice geometry and confinement produces flat bands with a non-trivial quantum geometry, without using spin-orbit coupling as in a previous effort~\cite{Scammell2019Tuning}.\ This is important as flat bands with a non trivial quantum geometry, may be able to produce a superfluid weight when electron-electron interactions are included~\cite{Peotta2015}.\ Actually, there has been some experimental progresses on fabricating high quality SL with clear evidence of mini-bands~\cite{Wang2017Observation,Du2021Observation,FranchinaVergel2021,wang2024,Wang2023Formation} and some similar proposals to induce unconventional superconductivity~\cite{Ingham_2020,Ingham_doped}.\ For that reason, here we also explore the effects of including electrostatic interactions between electrons.

In this work we consider a two-dimensional electron gas placed on top of a patterned gate to derive exact analytical solutions describing the effects of a SL potential with square and triangular geometries. By providing a comprehensive phase diagram, we map the parameter space governing the emergence of narrow bands and pseudo-Landau levels.\ Additionally, by representing the Hamiltonian in a plane-wave basis, we track the evolution of the band structure as a function of the SL strength.\ By modifying the symmetry of the SL potential we induce a non-trivial quantum geometry.\ Finally, by incorporating self-consistent Hartree screening, we reveal a competition between the SL potential strength and the screening effects.

The layout of this work is the following: in Sec.~\ref{Sec:Model} we present the model of such structures and its electronic properties are discussed in Sec.~\ref{Sec:Flat}.\ Two important  examples, the rectangular and square lattices are presented in Sec.~\ref{Sec:Examples} while the triangular SL is presented in Sec.~\ref{Sec:Hexa}. In Sec.~\ref{Fourier} we introduce the Fourier representation of the SL potential.\ In Sec.~\ref{topology} we analyze the quantum geometry and in Sec.~\ref{electrostatic} we describe the electrostatic interactions.\ Finally, the conclusions are given in Sec.~\ref{Sec:Conclusions}.

\section{Model}\label{Sec:Model}
\begin{figure}[t]
    \centering
     \includegraphics[scale=0.95]{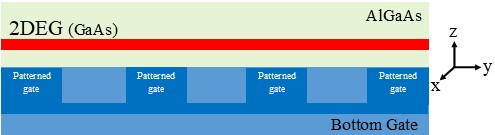}
    \caption{Schematics of a patterned gate acting on a GaAs heterostructure.\ The bottom gate modulates the strength of the potential and the patterned gate modulates the local charge density.}
    \label{fig:Figure1}
\end{figure}

We consider a two-dimensional GaAs heterostructure subjected to the effect of a patterned gate~\cite{wang2024Tuning}.\ As shown in Fig.~\ref{fig:Figure1}, the metallic patterned gate is placed at the bottom of a semiconductor.\ The voltage on the bottom gate creates anti-dots in the two-dimensional electron gas under the perforations in the patterned gate.\ The combination of both gates creates an electrostatic potential $U(\mathbf{r},z)$, where $\mathbf{r}=(x,y)$ is the position in a plane at height $z$. Such potential satisfies the Laplace's equation~\cite{Tkachenko2014Effects}.\ Because the patterning is assumed to be periodic, we also have $U(\mathbf{r},z) = U(\mathbf{r} + \mathbf{L}_j,z)$, with $\mathbf{L}_j$ being a primitive lattice vector defining the periodicity along the $j$-direction. Associated to this lattice, we define the reciprocal vectors $ \mathbf{G} = \mathbf{G}_{mn} = m \mathbf{G}_1 + n \mathbf{G}_2$,  given in terms of a given basis $\{ \mathbf{G}_1, \mathbf{G}_2\}$ determined by the SL grid as usual. 

By setting our coordinate system at the top of the perforated gate, the corresponding potential is determined by the solution of the Laplace's equation with a boundary condition determined by $U(\mathbf{r}, z=0)$, which tracks the shape of the pattern. This potential at $z=0$ is a constant whenever $\mathbf{r}$ is in a region where the metal is present and otherwise has a different value for $\mathbf{r}$ in other regions (see~Fig.~\ref{fig:Figure1}).

Taken into account the periodicity, the geometry and the boundary conditions, $U(\mathbf{r},z)$ can be expanded using  a proper and complete set of basis functions suitable for a cilindrical geometry. In this case, such expansion is made from the exponentials $e^{\pm G z}$ in $z$ and the complex exponentials $e^{i\mathbf{G}\cdot \mathbf{r}}$.\ Here the harmonics are labeled by the reciprocal lattice vectors $\mathbf{G}$. Therefore, up to a constant, the most general gate potential can be written as~\cite{Tkachenko2014Effects},
\begin{equation}
    U (\mathbf{r},z) = \underbrace{-e \mathcal{E} z }_{\text{boundary condition, } z \rightarrow \infty} + \sum_{\mathbf{G}  \in \Lambda^*} \alpha_{\mathbf{G}} e^{-G z} e^{i\mathbf{G} \cdot \mathbf{r}},
\end{equation}
where  $\Lambda^*$ is the reciprocal lattice, $\mathcal{E}$ is a uniform electric field generated by the gate and $\alpha_{\mathbf{G}}$ is the corresponding Fourier coefficient. Thus the potential is rewritten as
\begin{equation}
    U (\mathbf{r},z) =  -e \mathcal{E} z + 2\sum_{mn} \alpha_{mn} e^{-G_{mn} z} \cos(\mathbf{G}_{mn} \cdot \mathbf{r}+\phi_{mn}),
    \label{eq: Upote}
\end{equation}
where $\phi_{mn}$ is a phase that arises from the complex nature of the Fourier coefficients $\alpha_{mn}$, which depends on the symmetries of the SL potential and $G_{mn}=|\mathbf{G}_{mn}|$.
The Fourier coefficients are found by projecting the boundary condition at $z=0$ onto the basis functions inside the unitary cell $S$,
\begin{equation}
    \alpha_{\mathbf{G}}= \frac{1}{S}\int_{S} d^{2}{\mathbf{r}} U(\mathbf{r},0)e^{-i\mathbf{G} \cdot \mathbf{r}} 
\end{equation} 
For a triangular lattice with lattice constant $L$ made by drilling circular holes of radius $D$ in the metal, it is easy to prove that for example $\alpha_{10}$ is proportional to $J_1(2\pi D/\sqrt{3}L)$, where $J_1(x)$ is the Bessel function. For square holes in a square lattice, the harmonics are basically given by the Fourier expansion of a square wave.

To reproduce the potential at or near the perforated metallic plate, the Fourier expansion must contain high order components. Yet here we are only interested in the potential at the 2DEG layer. Therein, the first term in Eq.~(\ref{eq: Upote}) is constant.\ Moreover, as the 2DEG is away from the metallic gates, the factor $e^{-G_{mn} z}$ makes the high-frequency harmonics to decay rapidly, and therefore we can neglect all terms with $G_{mn} > \text{min}(|\mathbf{G}_{mn}|)$ and thus retain the first harmonics.\ Finally, by considering a symmetric lattice we can see that $\alpha_{mn}$ is the same for the first harmonics, let's call them $W$. Therefore, inside the 2DEG layer and away from the metallic SL we get the effective potential,
\begin{equation}
\label{eq: periodic potential}
    U (\mathbf{r}) \approx 2W \sum_{|m|,|n| = 1} \cos(\mathbf{G}_{mn} \cdot \mathbf{r}+\phi_{mn}).
\end{equation}
where the dependence on $z$ was now eliminated.
This results in a simplified model governed by an effective Hamiltonian in two dimensions which depends only on a single parameter $W$ representing the amplitude of the applied potential~\footnote{We note that an exact solution for a patterned gate acting in a 2DEG strongly depends on the geometry of the device and a full self-consistent Poisson-Schrodinger is then required~\cite{zhan2024designing}}, yet capturing the essential physics~\cite{Guinea2010Band,zhan2024designing}.\ Therefore, under our simplified model the total Hamiltonian is written as
\begin{equation}
\label{eq: total hamiltonian}
H  =\frac{p^2}{2 m^*}+U(\mathbf{r}),
\end{equation}
where $m^*$ is the electron effective mass in GaAs. By construction the SL potential is a smooth periodic function such that $U(\mathbf{r}+\mathbf{L}_1)=U (\mathbf{r}+\mathbf{L}_2)=U (\mathbf{r})$, with $\mathbf{L}_1$ and $\mathbf{L}_2$ primitive lattice vectors.\ For a triangular (or square) SL, the lattice period is given by $L_m = |\mathbf{L}_1|=|\mathbf{L}_2|$.\ The reciprocal lattice vectors satisfy $\mathbf{G}_i\cdot \mathbf{L}_j=2\pi \delta_{ij}$.\ For a triangular SL, the corners of the SL Brillouin Zone (sBZ) are given by $\mathbf{K} = (1/3)(\mathbf{G}_{1}+2\mathbf{G}_{2})$ and $\mathbf{K}' = (1/3)(2\mathbf{G}_{1}+\mathbf{G}_{2})$.\ In the considered geometries, the SL potential is given by
\begin{equation}
\label{eq: reg potential}
    U(\mathbf{r})  =2 W\sum_j \cos \left(\mathbf{G}_j \cdot \mathbf{r}+\phi_j \right),
\end{equation}
where $\phi_j$ is a phase on each direction defined by $\mathbf{G}_j $.\ In this work we will consider the case $\phi_j=\phi$ for all $j$.\ It is important to note that for the rectangular and square lattices, the phase can always be removed by a simple translation while this is not the case for
the triangular lattice.\ To see this consider the transformation $\mathbf{r'}=\mathbf{r}-\mathbf{r}_0$ such that $\mathbf{G}_1 \cdot \mathbf{r}_0=-\phi$ and $\mathbf{G}_2 \cdot \mathbf{r}_0=-\phi$. It follows that
\begin{equation}
\begin{aligned}
    U(\mathbf{r'})=2 W [\cos \left(\mathbf{G}_1 \cdot \mathbf{r'}\right)+\cos \left(\mathbf{G}_2 \cdot \mathbf{r'}\right) \\
   +\cos\left(\mathbf{G}_3 \cdot \mathbf{r'}+(\phi+\mathbf{G}_3)\cdot \mathbf{r}_0\right)]
\end{aligned}
\end{equation}
The last term is zero for the square and rectangular lattices while for the triangular lattice still depends on $\phi$, as from $\mathbf{G}_1+\mathbf{G}_2+\mathbf{G}_3=0 $ we obtain
\begin{equation}
\begin{aligned}
    U(\mathbf{r'})&=2 W [\cos \left(\mathbf{G}_1 \cdot \mathbf{r'}\right)+\cos \left(\mathbf{G}_2 \cdot \mathbf{r'}\right) \\ 
    &+\cos\left(\mathbf{G}_3 \cdot \mathbf{r'}+3\phi \right)].
\end{aligned}
\label{eq: Ugeneral}
\end{equation}
Thus, the relative phase of the third component can not be ruled out unless $\phi=2\pi s/3$ with $s=0,1,2$.\ Since a sign change in $W$ is equivalent to a phase shift of $\phi=\pi$, this leads to the following consequence: while a sign change in $W$ for square or rectangular lattices is merely equivalent to a shift of the origin, this is not the case for the triangular lattice.\ As a result, the spectra of square and rectangular lattices are expected to remain invariant under a sign change of $W$, whereas for the triangular lattice, a different spectrum is anticipated. 

We note that the simple form of the SL potential in Eq.~\eqref{eq: reg potential} has been successfully used to describe the scalar contributions of hexagonal boron nitride (hBN) substrates acting in graphene monolayers~\cite{Wallbank2013Generic,SanJose2014Spontaneous,Jung2015Origin,zhan2024designing} and graphene bilayers~\cite{ghorashi2023multilayer,zeng2024gate}.\ In the following section we will discuss the electronic properties derived from such a SL potential acting on a bidimensional electron gas.

\section{Electronic properties: quasi-free electrons, flat bands and pseudo-Landau levels}\label{Sec:Flat}

The electron dynamics is described by the Schr\"{o}dinger equation
\begin{equation}
    -\frac{\hbar^{2}}{2 m^*}\nabla^2\psi(\mathbf{r})+U(\mathbf{r})\psi(\mathbf{r})=E\psi(\mathbf{r}),
    \label{eq:Schro}
\end{equation}
where the potential is given by Eq.~\eqref{eq: reg potential}.\ Let us discuss how the solutions to Eq.~(\ref{eq:Schro}) contain two limiting cases depending on the dimensional parameter as
\begin{equation}
    q^{*}_j=\frac{|W|}{\hbar^{2} |\mathbf{G}_{j}|^{2}/2m^{*}},
\end{equation}
where $j=1,2$.\ As detailed in  the following subsections and in Appendix~\ref{Sec:AppExpnasion}, this parameter measures the ratio between the maximal potential energy  and maximal kinetic energy for electrons in the quasi-free electron approximation along the $j$ direction.\ While for some energies the solutions can be described within the free electron approximation and resemble the potential shape, for others, the system behaves as a set of nearly isolated, deformed quantum harmonic oscillator potentials centered at the minima of  $U(\mathbf{r})$. Mathematically, this occurs because, in the quantum oscillator limit, the wave functions require large gradient envelopes to compensate for the strong potential energy.\ Although later on this will be explained in detail, in a neighborhood of the $U(\mathbf{r})$ minima, denoted generically as $\mathbf{r}_{m}$, the potential can be approximated as
\begin{equation}
    U(\mathbf{r}) \approx U(\mathbf{r}_{m})+\frac{1}{2}\mathbf{\delta}^{T} \mathcal{D}(\mathbf{r}_{m}) \mathbf{\delta}
    \label{eqn:Pot}
\end{equation}
where $\mathbf{\delta}^{T}=\mathbf{r}-\mathbf{r}_m$. $\mathcal{D}(\mathbf{r}_{m})$ is the Hessian matrix evaluated at such extremal points, with elements
\begin{equation}
    \mathcal{D}_{\mu \nu}(\mathbf{r}_{m})=  -2W \sum_{j} G_j^{\mu}G_j^{\nu}e^{i\mathbf{G}_j\cdot \mathbf{r}_m}.
\end{equation}
where $G_j^{\mu}$ denotes the $\mu=x,y$ components of the vector $\mathbf{G}_j$.\ Due to the quadratic nature of Eq.~\eqref{eqn:Pot}, pseudo-Landau levels appear within each potential minimum, producing flat bands at low energies.\ We emphasize that the bands arise because the system is always periodic, even in the strong confinement limit.\ Consequently, we advance that the spectrum differs from that of a set of isolated quantum dots, whose spectrum consists of discrete levels~\cite{Macucci1992Electronic}.\ Below, we discuss some examples of this behavior.

\subsection{Rectangular and Square Lattices}
\label{Sec:Examples}

Let's first consider a rectangular SL unit cell with dimensions $L_1$ and $L_2$ along the $x$ and $y$ axes respectively.\ We define the reciprocal vectors $\mathbf{G}_{\mathbf{1}} = (2\pi/L_1)(1,0)$ and $\mathbf{G}_{\mathbf{2}} =(2\pi/L_2) (0,1)$. The SL induced potential reads as
\begin{equation}
    U(\mathbf{r}) =2 W\left[\cos \left(G^x_1 x \right)+\cos \left(G^y_2 y \right) \right].
\end{equation}
The corresponding Schr\"{o}dinger equation $H\psi(\boldsymbol{r})=E\psi(\boldsymbol{r})$, with $E$ the energy and $\psi(\boldsymbol{r})$ the wave function, is separable into two ordinary second order differential equations by proposing a solution of the type $\psi(\boldsymbol{r})=X(x)Y(y)$, as
\begin{equation}
\label{eq: x}
   \frac{d^{2}X(x)}{dx^2}+\frac{2m^*}{\hbar^{2}}(E_1-2W \cos(G^x_1 x))X(x)=0,
\end{equation}
\begin{equation}
\label{eq: y}
    \frac{d^{2}Y(y)}{dy^2}+\frac{2m^*}{\hbar^{2}}(E_2-2W \cos(G^y_2 y))Y(y)=0,
\end{equation}
where $E=E_1+E_2$.\ The previous equations are solvable using standard band theory techniques, i.e., by proposing  Bloch wave solutions.\ In that case, $E_1$ and $E_2$ are functions of the wavevector $\mathbf{k}=(k_x,k_y)$. The details of this approach are presented in Sec.~\ref{Fourier}. The resulting spectrum is seen in Fig.~\ref{fig:Bandcharge} for the cases $W=0.25,1.0$ and $W=3.0$ meV.\ Notice how the lower bands become flat as the confinement increases. To understand the details of how this happens, here we prefer to use a slightly different theoretical approach.\ This allows us to find analytical solutions for the wavefunctions.\ The trade-off is that the energy dispersion $E(\mathbf{k})$ is not straightforward.\ Nonetheless, the corresponding $\mathbf{k}$-values for the band edges are particularly easy to determine as they correspond to high-symmetry points in reciprocal space.\ Therefore, here we observe that Eq.~\eqref{eq: x} and Eq.~\eqref{eq: y} are Mathieu equations with the generic form 
\begin{equation}
    \frac{d^2\theta(t)}{dt^2}+[a-2q\cos(2t)]\theta(t)=0,
\end{equation}
that are well known to describe a classical parametric driven pendulum, where $\theta(t)$ is the angular coordinate of the pendulum at time $t$. After a change of variables, Eq.~\eqref{eq: x} and Eq.~\eqref{eq: y} are written as a pair of Mathieu equation with a set of different $a,q$ parameters
\begin{equation}
    a_j=8\frac{m^{*}E_j}{\hbar^{2}G_j^{2}}, \ \ \ \ q_j=16\frac{m^{*}|W|}{\hbar^{2}G_j^{2}},
    \label{eq:aqdef} 
\end{equation}
where $j=1$ is used for Eq.~\eqref{eq: x} and $j=2$ for Eq.~\eqref{eq: y}.\ Observe that $q_j$ is defined in Eq. (\ref{eq:aqdef}) using the absolute value and the reason is the symmetry of the Matthieu equation with respect to $q$.\ Notice also that $q_j$ is a refined version of the parameter $q^{*}_j$ and basically represents the ratio between potential and kinetic quasi free particle energy. The solutions to the Mathieu equation are found using Floquet theory.\ They are linear combinations of the Mathieu cosine and sine functions, $\operatorname{ce}(a,q,t)$ and  $\operatorname{sn}(a,q,t)$, respectively  
\begin{subequations} \label{eq: MatEigensol} 
\begin{equation}
\begin{aligned}
    X(x)=&A_{ce}^{x}\operatorname{ce}(a_1,q_1,G^x_1x/2) \\
    &+A_{sn}^{x}\operatorname{sn}(a_1,q_1,G^x_1x/2),
\end{aligned}
\end{equation}
\begin{equation}
\begin{aligned}
    Y(y)=&A_{ce}^{y}\operatorname{ce}(a_2,q_2,G^y_2y/2)\\
    &+A_{sn}^{y}\operatorname{sn}(a_2,q_2,G^y_2y/2),
\end{aligned}
\end{equation}
\end{subequations} 
which are equivalent to the Bloch solutions.\ Here $A_{ce}^{x}$, $A_{sn}^{x}$, $A_{ce}^{y}$ and $A_{sn}^{y}$ are constants that depend on the boundary conditions.\ For example, symmetric solutions $X(-x)=X(x)$ require $A_{ce}^{x}=1$ and $A_{sn}^{x}=0$ while antisymmetric ones $X(-x)=-X(x)$
requires $A_{ce}^{x}=0$ and $A_{sn}^{x}=1$.\ The same is true for the $y$ axis Mathieu equation. 

\begin{figure}[t]
    \centering
     \includegraphics[scale=0.80]{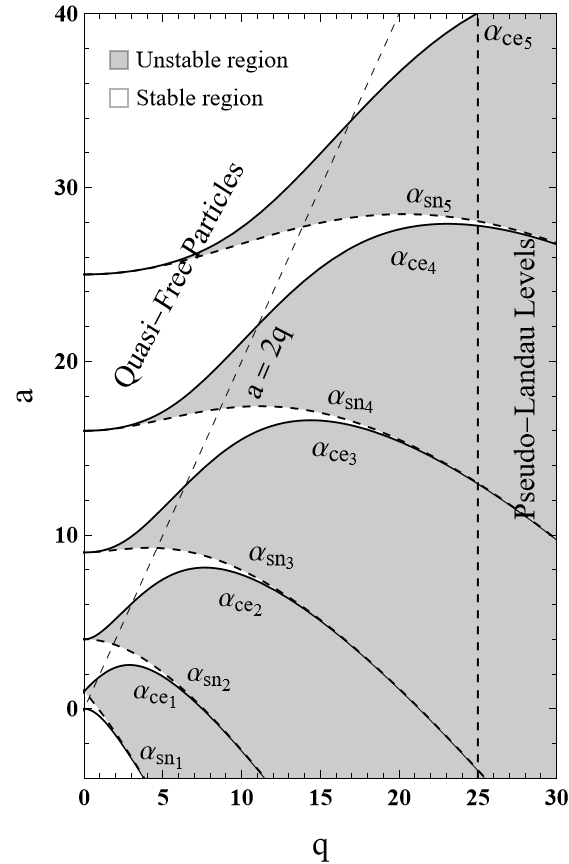}
    \caption{Mathieu equation stability chart, here corresponding to the eigenenergies of the square lattice on each direction.\ Stable regions correspond to bands while unstable regions are spectral gaps of the system. The solid and dashed curves, $\alpha_{ce}(q)$ and $\alpha_{sn}(q)$, stand for the eigenvalues of Mathieu's Floquet matrix corresponding to the eigenvectors (Mathieu functions) $\operatorname{ce}(a,q,t)$ and $\operatorname{sn}(a,q,t)$.\ The two different limiting regimes are highlighted with the transition line $a=2q$, where  the amplitude of the driven perturbation matches the undriven harmonic oscillator parameter.\ The vertical dotted line indicates how the spectrum is read for a given $W$, showing pseudo Landau levels at low energies.\ Only the $q\ge 0$ region is presented as the chart is the same for $q \le 0$. }
    \label{fig:stability}
\end{figure}

\subsection{Interpretation of the Stability Chart}

The Mathieu equation presents  bounded periodic solutions only for certain combinations of the parameters $a$ and $q$~\cite{mclachlan1964theory}.\ The stability chart seen in Fig.~\ref{fig:stability} contains such information.\ Unstable solutions are associated with resonances between the fundamental frequency of the pendulum and the driving force $2q\cos(2t)$.\ For $q=0$, the undriven harmonic oscillator is recovered corresponding to the undriven pendulum.\ This is equivalent to turning off the SL potential.\ Along the line where $a=2q$, the amplitude of the cosine matches the undriven harmonic oscillator parameter.\ The stability condition in the Mathieu chart translates into allowed energies, i.e., unstable regions are spectral gaps while stable regions corresponds to the bands. 

Here, the role of the pendulum is played by the free-particle solutions while the SL potential provides a spatial driving akin to the temporal driving.\ As seen in Fig.~\ref{fig:stability}, for $q<<1$ the gaps are open  at $a\approx n^{2}$ with $n=1,2,3,...$, each corresponding to a resonance between the natural frequency of the pendulum and the frequency of the driving force.\ These gaps are translated here into a resonance between the spatial driving frequency $|G_j|$ and the wave vector component $k_j$. Physically, the gaps are open by diffraction as stationary waves are created by destructive or constructive interference whenever $k_j \approx n G_j$.

As we are dealing in our lattice problem with two equations, both $a_1,q_1$ and $a_2,q_2$  must lay in stable regions.\ Therefore, the Mathieu stability chart must be applied to both the $x$ and $y$ directions. Since $E=E_1+E_2$, the spectrum is degenerate as many combinations are possible for a single $E$. Also observe how $q$ is fixed by $W$ as indicated by a dotted vertical line in Fig.~\ref{fig:stability}.\ The spectrum is read for a fixed $W$ along such vertical line where each allowed value of $a$ gives its corresponding energy. Each tongue in the Mathieu chart corresponds to a Bloch's energy band, and within each tongue,  all possible $\mathbf{k}$ values in one direction are considered.\ The pure Mathieu cosine and sine functions correspond to each of the band edges seen in Fig.~\ref{fig:stability}, as they have periodic and antiperiodic boundary conditions.
Also, observe that in Eq.~\eqref{eq: MatEigensol}, for a given combination of $E_j$ and $W$ in Eq.~\eqref{eq:aqdef}, the charge density in real space appears to be obtained without requiring a Bloch representation.\ In fact, such a representation is embedded in the Mathieu functions and the stability chart, as both are calculated using a matrix determinant similar to Bloch's approach.\ However, this determinant can be analytically evaluated using a method developed by Whittaker~\cite{mclachlan1964theory}, enabling the derivation of a closed-form solution.

\subsection{Limiting Cases for the Rectangular Lattice}

\begin{figure*}[ht]
    \centering
     \includegraphics[scale=0.50]{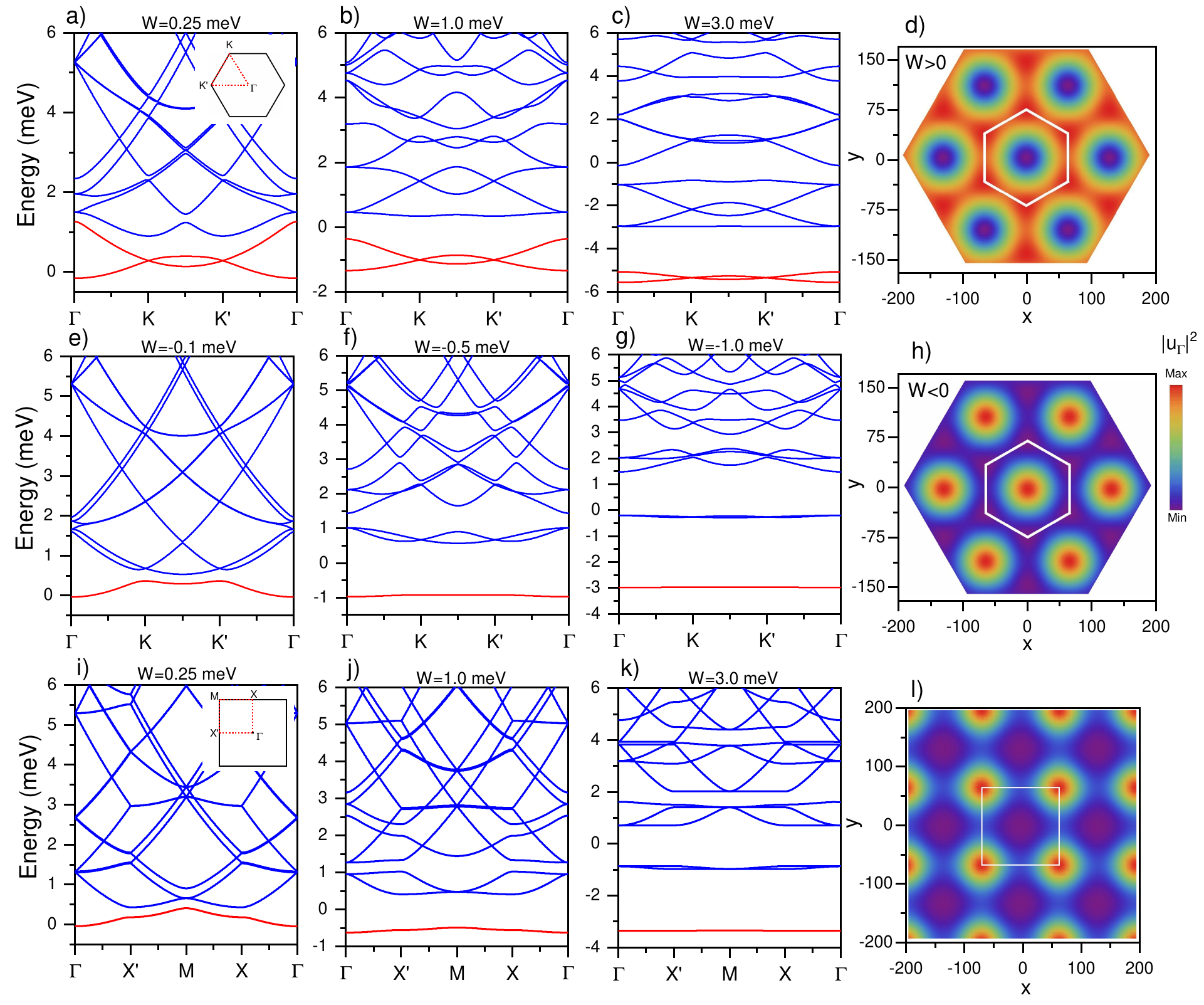}
    \caption{Electronic structure of GaAs subjected to a patterned SL potential.\ Top and middle rows are the bands with a triangular patterning and bottom row is for a square patterning.\ Red line are the lowest energy bands.\ Panels d), h) and i) are the electronic charge densities at the $\Gamma$ point for the corresponding patterning.\ Our parameters are such that $E_0=\hbar^2/2m^* = 0.56$ $eV \text{nm}^2$, $m^* =0.067 m_e$, with $m_e$ the electron mass and the SL length is set to $L=130$ nm.\ The high symmetry points and the evaluation paths are shown in a) and i).\ Observe how in c), g) and k) the lowest bands are almost flat reaching the Landau level limit.\ Also, the ground states in d), h) and i) resemble the $U(\mathbf{r})$ in agreement with Eq.~\eqref{eq: uGamma}.\ Notice the presence of Dirac cones in panels a), b) and c).}
    \label{fig:Bandcharge}
\end{figure*}

Now, let us discuss some limiting cases.\ Consider first the limit $2q_1>>a_1$ and $2q_2>>a_2$. In this case the driving force dominates.\ The solutions remain close to the minima of $\cos(2t)$. Therefore, the Mathieu equation reduces to the harmonic oscillator Schr\"{o}dinger’s equation and, locally, the corresponding wavefunctions are given by Landau level solutions \cite{WILKINSON2018}
\begin{equation}
    X(x)=c_{n_1}H_{n_1}(2^{-3/4}q_1^{1/4}G^x_1x)e^{-\frac{1}{2}\sqrt{2q_1}(\frac{G^x_1x}{2})^{2}},
    \label{eq:x}
\end{equation}
\begin{equation}
    Y(y)=c_{n_2}H_{n_2}(2^{-3/4}q_2^{1/4}G^y_2y)e^{-\frac{1}{2}\sqrt{2q_2}(\frac{G^y_2y}{2})^{2}},
    \label{eq:y}
\end{equation}
where $c_{n_1}, c_{n_2}$ are two constants and $H_n(x)$ is a $n$-degree Hermite polynomial.\ The allowed parameters are
\begin{equation}
    a_j=4\sqrt{q_j}\left(n_j+\frac{1}{2}\right)-2q_j
\end{equation}
with $n_j=0,1,2,3,...$.\ The resulting spectrum is akin to the Landau level spectrum
\begin{equation}
    E_{n_1,n_2}=\hbar \omega_1\left(n_1+\frac{1}{2}\right)+\hbar \omega_2\left(n_2+\frac{1}{2}\right)-8W,
\end{equation}
with frequencies
\begin{equation}
    \omega_j=2G_j\left(\frac{W}{m^{*}}\right)^{1/2}.
\end{equation}
Therefore, flat bands arise as the confinement increases due to the minima of the SL potential. The transition start to occur once $q_{1,2}>a_{1,2}/2$, i.e., for $E_1<16W$ and $E_2<16W$.  Notice that for the square lattice, $G^x_1=G^y_2$ and $q_1=q_2$, resulting in 
\begin{equation}
    E_{n_1,n_2}=\hbar \omega_c(n_1+n_2+1)-8W,
\end{equation}
where we defined an effective cyclotron frequency
\begin{equation}
    \omega_c \equiv 2 G^x_1\left(\frac{W}{m^{*}}\right)^{1/2}=\frac{4\pi}{L} \left( \frac{W}{m^{*}}\right)^{1/2}, 
\end{equation}
with $L\equiv L_1=L_2$. Such results are in agreement with the numerical spectrum seen in Fig.~\ref{fig:Bandcharge} for the cases $W=0.25,1.0$ and $W=3.0$ meV.\ Therein, the confinement is strong for the lower bands. It is observed either with a strong potential or a large lattice parameter, $L$.\ This partly explains why it has not yet been observed, as previous efforts have focused on smaller lattice parameters \cite{VonKlitzing1999}.

We remark here that Eq.~\eqref{eq:x} and Eq.~\eqref{eq:y} are local approximations to the analytical periodic solution given in terms of Mathieu functions.\ A global approximation around all minima of the potential is
\begin{equation}
    \psi(\mathbf{r}) \approx \sum_{m,n}X(x-mL_{1})Y(y-nL_2).
\end{equation}
It is seen from here that the coordinates $(mL_{1},nL_2)$ play the role of the guiding coordinates in the quantum Hall effect.\ Also we remark again that, as the Mathieu functions are periodic, the spectrum is always made from bands and not isolated levels as happens with quantum dots.\ This behavior is similar to that observed in magic angle twisted bilayer graphene where flat bands have periodic Gaussian cores with a small overlap between them~\cite{Elias_2023,Song2022Magic}.

Let us discuss the other limit.\ For $q_1=0$ and $q_2=0$ all values of $a_1$ and $a_2$ are allowed and correspond to a continuous spectrum, a result to be expected due to the absence of a SL potential.\ The weak SL modulation case corresponds to $2q_1<<a_1$ and $2q_2<<a_2$. Gaps are open at
\begin{equation}
    a_j \approx n_j^{2},
\label{eq:alimit}
\end{equation} 
corresponding to the resonances of a weak perturbed parametric pendulum.\ In this case, the spectral gaps, i.e. resonances, are due to the induced SL diffraction as can be readily confirmed by using Eq.~\eqref{eq:alimit} and Eq.~\eqref{eq:aqdef}, as
\begin{equation}
    E_j=\frac{\hbar^{2}}{2m^{*}}\left(\frac{n_1G^x_1}{2}\right)^{2}=\frac{1}{2m^{*}}\left(\frac{n_1P_1}{2}\right)^{2},
\end{equation}
showing that $n_1 G^x_1$ acts as a wave vector that opens a gap.\ As expected, in this limit the charge carriers behave as usual free particles under a weak periodic perturbation.\ We note that this weak coupling limit has been widely used to introduce triangular SL potentials in graphene heterostructures~\cite{Wallbank2013Generic,Mucha2013Heterostructures,Jung2017Moire,SanJose2014Electronic,SanJose2014Spontaneous,Jung2015Origin,zhan2024designing}. 

\subsection{Triangular Superlattice}
\label{Sec:Hexa}
 
Now we will study a triangular SL, akin to the moiré effective potential that appears in twisted bilayer graphene~\cite{SanJose2012NonAbelian}.\ In Fig.~\ref{fig:Bandcharge}, the top and middle rows are the bands for a triangular patterning. Notice the presence of Dirac cones in panels of Fig.~\ref{fig:Bandcharge}a)-c) suggesting the presence of interesting topological properties when compared with the square lattice.

Some interesting features arise when the patterning is triangular. For $\phi_j = 0$ and $W > 0$ in Eq.\eqref{eq: reg potential}, the SL potential is repulsive at the center of the unit cell and attractive at its edges.\ This configuration gives rise to an effective honeycomb SL potential. Conversely, when $W < 0$, the SL potential becomes attractive at the center and repulsive at the edges, resulting in an effective triangular SL potential. The ground-state charge distribution localizes in the attractive regions, forming a lattice structure. As illustrated in Fig.~\ref{fig:Bandcharge}d) and Fig.~\ref{fig:Bandcharge}h), this allows the charge density maxima of the ground state to transition from a honeycomb to a triangular lattice by tuning the value of $W$. This symmetry change directly reflects the shift in the lattice defined by the minima $\mathbf{r}_m$ of the potential $U(\mathbf{r})$. As explained in Sec.~\ref{Sec:Model}, for $W<0$ the sign can be absorbed into the phase $\phi=\pi$ and the potential maps into
\begin{equation}
    \begin{aligned}
    U(\mathbf{r'})&=2 |W| [\cos \left(\mathbf{G}_1 \cdot \mathbf{r'}\right)+\cos \left(\mathbf{G}_2 \cdot \mathbf{r'}\right) \\ 
    &-\cos\left(\mathbf{G}_3 \cdot \mathbf{r'}\right)].
\end{aligned}
\end{equation}
The sign in the last term is responsible for the change of the minima positions from honeycomb to triangular.\ In other words, for $W>0$ the potential minima form a triangular lattice and the maxima form a honeycomb lattice. 

As the honeycomb lattice has a bipartite structure while the triangular lattice does not, this difference carries a significant implication for the low-energy band structure.\ This is seen when comparing Fig.~\ref{fig:Bandcharge}a) and Fig.~\ref{fig:Bandcharge}e).\ The honeycomb lattice ($W>0$) supports a doublet band~\cite{Wang2017Observation}, while the triangular lattice ($W<0$) only contains a singlet band~\cite{Wang2023Formation}.\ The doublet band appears in all bipartite lattices, as one band can be obtained from the other by reversing the sign of the wave function on one sublattice.\ Alternatively, this can be understood in terms of the honeycomb lattice, which hosts two Wannier orbitals per unit cell, each localized on a different sublattice.\ A mass term can then be introduced by making the two energy minima inequivalent, leading to a gap at the crossing points (See Sec.~\ref{topology}).\ In Appendix~\ref{Sec:AppLargeW}, we exemplify such effect in the low energy bands of a honeycomb SL by considering a sine potential which is gradually mixed with the cosine potential, resulting in an effective mass term that opens a gap.\ Additionally, electronic frustration plays a crucial role when comparing bipartite and non-bipartite lattices.\ The distance between the minima is also important and this  results in quite different ground state energies when comparing Fig.~\ref{fig:Bandcharge}d) and Fig.~\ref{fig:Bandcharge}h).

We now derive some analytical expressions. Let us start with a triangular SL with spacing $L_m$.\ The SL is generated by the basis
\begin{align*}
    \textbf{G}_1 &= \frac{2\pi}{L}(1,-\frac{1}{\sqrt{3}}), \\
    \textbf{G}_2 &= \frac{4\pi}{L\sqrt{3}}(0,1), \\
    \textbf{G}_{3} &=-(\textbf{G}_2 + \textbf{G}_1),
\end{align*}
with $\left|\mathbf{G}_{\mathbf{j}}\right|\equiv G=4 \pi / \sqrt{3} L$. 
The SL, considering for simplicity $\phi=0$, is written as
\begin{equation}
     U(\mathbf{r})  =2 W \sum_{j=1}^{3} \cos \left(\mathbf{G}_j \cdot \mathbf{r}\right).
\label{eq: SLPotential}     
\end{equation}
For $W>0$, the minimum of the potential is $U=-3 W$ and the maximum is $U=6 W$, such that the potential amplitude is $9W$. Let us simplify the previous expression by defining
$u=\boldsymbol{G}_1\cdot \boldsymbol{r}, v=\boldsymbol{G}_2\cdot \boldsymbol{r}, w= \boldsymbol{G}_3\cdot \boldsymbol{r}$. Then we use triangular coordinates such that
\begin{align*}
    \zeta &= u+v, \\
    \eta &= u-v.
\end{align*}
In triangular coordinates, the operator $\frac{p^2}{2m^*}$ is written as
\begin{equation}
    -\frac{\hbar^2G^2}{2m^*} \left( \frac{\partial^2}{\partial \zeta^2} + 3 \frac{\partial^2}{\partial \eta^2} \right).
\end{equation}
The potential is written as
\begin{equation}
U(\zeta,\eta)=2W\left[\cos\frac{\zeta+\eta}{2} + \cos \frac{\zeta - \eta}{2}+ \cos \zeta\right],
\end{equation}
or
\begin{equation}
U(\zeta,\eta)=2W\left[\cos\zeta+2\cos \frac{\eta}{2}\cos \frac{\zeta}{2} \right].
\end{equation}
The advantage of the triangular coordinates is that  the unitary cell is transformed into a rectangular domain defined by $0\le\eta \le 2\pi$  and $0\le\zeta \le 2\pi$.\ Also, such coordinates reflect in a natural way the symmetries of the triangular lattice unitary cell.\ In the minima of the potential, there is a strong confinement.  
Consider the case $W<0$.\ The strong confinement occurs at the minima of the potential, i.e. around $\mathbf{r
}_0=0$ and it implies that $\eta<<1$ and $\zeta<<1$, from where
\begin{equation}
\begin{aligned}
    E\psi &=-\frac{\hbar^2G^2}{2 m^*}\left(\frac{\partial}{\partial \zeta^{2}}+ 3\frac{\partial}{\partial \eta^{2}}\right)\psi \\
    &+2 |W| \left[\frac{3\zeta ^2}{4}+\frac{\eta^2}{4}-3 \right]\psi.
\end{aligned}
\end{equation}
 The previous equation can be separated resulting in two uncoupled quantum harmonic oscillators, i.e.,  using $\psi(\zeta,\eta) = \gamma(\zeta) \chi(\eta)$ we have
\begin{align}
    -\frac{\hbar^2G^2}{2m^*} \frac{\partial^2\gamma}{\partial \zeta^2}  + \frac{3|W|}{2} \zeta^2 \gamma  &= E_\zeta \gamma,  \\
    -\frac{3\hbar^2G^2}{2m^*} \frac{\partial^2 \chi}{\partial \eta^2}  +\frac{|W|}{2} \eta^2 \chi  &= E_\eta \chi,
\end{align}
then we can write the total energy as a sum of harmonic oscillator energies
\begin{equation}
    E=\hbar \omega_1\left(n_1+\frac{1}{2}\right)+\hbar \omega_2\left(n_2+\frac{1}{2}\right)+E_0,
\end{equation}
with
\begin{align}
    \omega_1 &= G\left(\frac{ 3|W|}{m^*}\right)^{\frac{1}{2}} = \frac{4 \pi}{ \sqrt{3} L}\left(\frac{3|W|}{m^*}\right)^{\frac{1}{2}},  \\
    \omega_2&=\omega_1, \\
    E_0 &= -6|W|.
\end{align}
Notice again that although the original problem was not radial symmetric, the local confinement allows to have an effective radial symmetric problem.\ Also, observe that the effective potential can be written as
\begin{equation}
    U_{\zeta,\eta}=-2|W|\left[2\cos \frac{\zeta}{2} \left(\cos \frac{\zeta}{2}+\cos \frac{\eta}{2}\right)-1\right],
\end{equation}
and
\begin{equation}
    U_{\zeta,\eta}=-2|W|\left[2\sum_{s=0} \frac{(-1)^{s}}{2s!}\left(\frac{\zeta}{2}\right)^{2s}\left(\cos \frac{\zeta}{2}+\cos \frac{\eta}{2}\right)-1\right].
    \label{eq: Ueff}
\end{equation}
For $s=0$ we recover a rectangular case, studied in the previous section.\ Higher orders in $s$ increases the confinement in the $\zeta$ direction.\ A similar analysis can be made for $W>0$ considering that the minima of the potential are no longer at  $\mathbf{r
}_0=0$.

\begin{figure*}[ht]
\centering
\includegraphics[width=0.9 \textwidth]{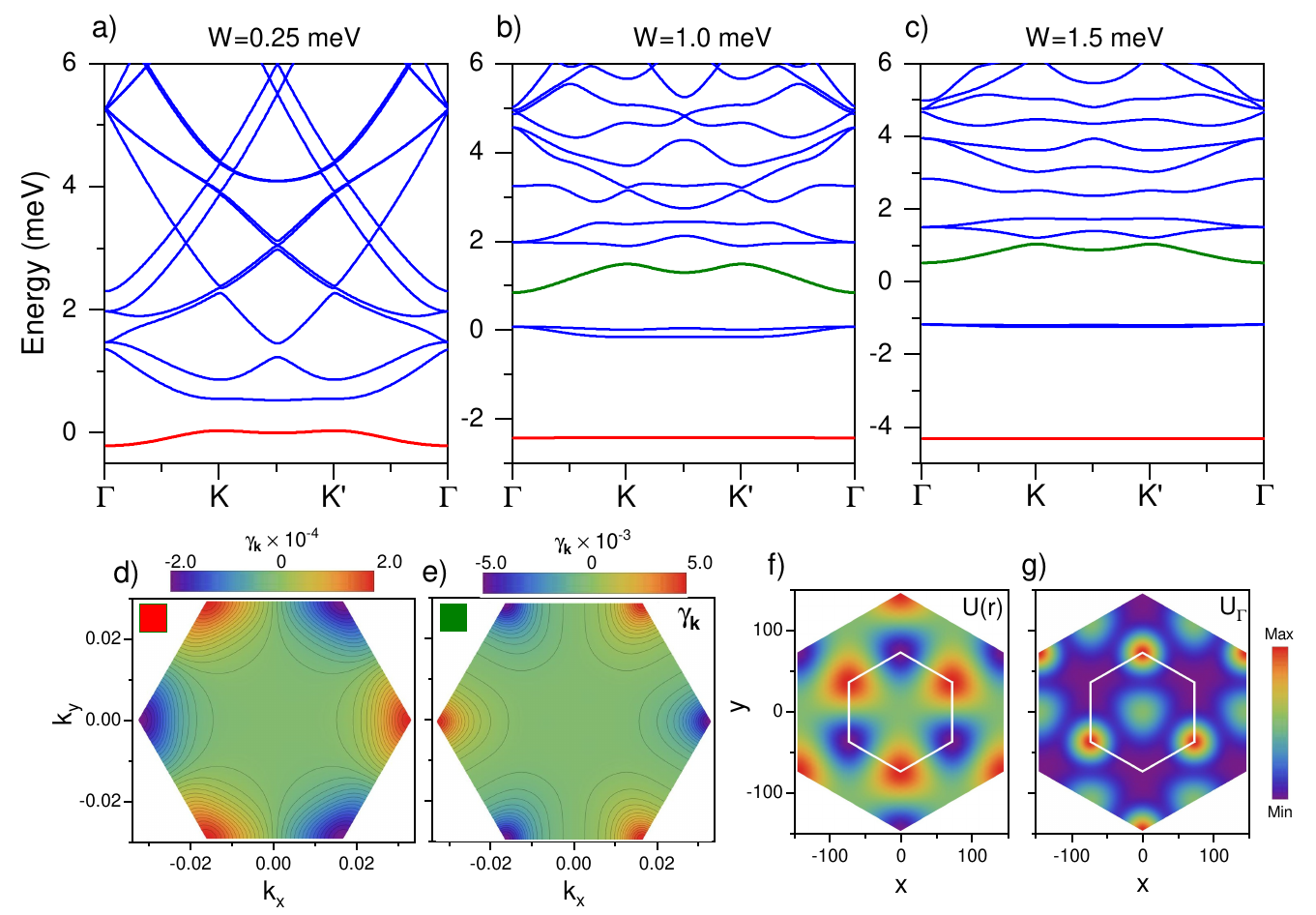}
\caption{Electronic structure of a 2DEGs subjected to an antisymmetric scalar SL potential with (a) $W = 0.25$ meV, (b) $W = 1.0$ meV, and (c) $W = 1.5$ meV. Panel (d) shows the local Berry phase corresponding to the bands highlighted in red in each panel, while panel (e) shows the local Berry phase for the bands highlighted in green. Panel (f) displays the antisymmetric SL potential, and panel (g) shows the charge density at the $\Gamma$ point for the low-energy band in (a). The white hexagon indicates the real-space unit cell.}
\label{fig:BandTopo}
\end{figure*}
\section{Nature of the electronic states} \label{nature}
In twisted bilayer graphene, it is known that at the first magic angle, the electronic density of the flat-band states reproduces the shape of the potential~\cite{SanJose2012NonAbelian}.\ For higher-order angles, these states take on a Gaussian profile with a pronounced tail~\cite{Navarro2022Why}.\ More recently, it has been shown that the flat-band states are topological solitons that resemble Gaussians as a first approximation, a result similar to the heavy fermion model~\cite{Song2022Magic}. A similar phenomenology is observed here.  Due to the Bloch's theorem, the electronic wave functions have the form
\begin{equation}
\psi_{\mathbf{k}}(\mathbf{r})=e^{\mathbf{k} \cdot \mathbf{r}}u_{\mathbf{k}}(\mathbf{r}),
    \label{eq:Psi}
\end{equation}
For the quasi-free particle $q_j^{*}<<1$ limit, in Appendix~\ref{Sec:AppExpnasion} we prove that the ground state  has the form
\begin{equation}
\psi_{\Gamma}(\textbf{r}) \approx C\left(1+\frac{2m^*}{\hbar^2|G|^2}(W-U(\textbf{r})) \right),
\label{eq: uGamma}
\end{equation}
where $C$ is a normalization constant and we assumed that $|\mathbf{G}_j|=G$ for all $j$. From Eq.~\eqref{eq: uGamma} we conclude that the ground state electronic density is proportional to the potential.\ Also notice how a change of sign in $W$ is reflected in a change of the electronic density location.\ Figure~\ref{fig:Bandcharge}d), Fig.~\ref{fig:Bandcharge}h) and Fig.~\ref{fig:Bandcharge}i) confirm the result given by Eq.~\eqref{eq: uGamma}, as the charge density at the $\Gamma$ point tracks the  potential with inverted sign. Such situation is akin to what happens for the first magic angle in twisted bilayer graphene~\cite{RademakerCharge2018}.\ In the case of strong confinement, the Landau level limit results as a boundary layer limit of the Schr{\"o}dinger equation as shown in the Appendix~\ref{Sec:AppExpnasion}.

\section{Fourier Space Representation}\label{Fourier}
We now consider a general form by using a Fourier representation such that we expand in plane waves the SL potential.\ This procedure is similar to that introduced in graphene monolayers with a triangular SL~\cite{Guinea2010Band}.\ As before, we define a pair of reciprocal lattice vectors, $\textbf{G}_1$ and $ \textbf{G}_2 $ defined by the SL in the folded Brillouin zone. The momentum $\textbf{q}$ of the 2D electron gas is written into a momentum $\textbf{k}$ within the boundaries of the sBZ and a contribution from the SL.\ We define the momentum $\textbf{k}$ inside the sBZ such that
\begin{equation}
\textbf{q}_{mn} = \textbf{k} + (m,n)\cdot (\textbf{G}_1, \textbf{G}_2) \equiv \textbf{k} + \textbf{G}_{mn},
\label{eq: Gmn}
\end{equation}
where $\textbf{G}_{mn}=m\textbf{G}_1+n\textbf{G}_2$ with $m,n$ integers. For a triangular SL, each $\textbf{G}_{mn}$ vector has six nearest neighbors, where the $\mathbf{G}_{mn}$ vectors with modulus $|\mathbf{G}_1|$, generate the so-called first harmonic functions~\cite{Wallbank2013Generic}.\ Successive harmonics are further apart from the $\textbf{G}=0$ origin.\ By considering the SL potential as a perturbation, the low-energy electronic structure in a plane wave basis is written as
\begin{align}
H & = H_0(\mathbf{k}) + U(r), 
\label{eq: hamprincipal}
\end{align}
where $H_0$ is the Hamiltonian of a 2D electron gas with matrix elements
\begin{align}
\left[H_{0}(\mathbf{k})\right]_{mn}=\frac{\hbar^2}{2m^{\ast}}\left|\mathbf{k}+\mathbf{G}_{mn}\right|^{2}.
\end{align}
The potential $U$ has matrix elements 
\begin{align}
    \left[U\right]_{mn,m'n'}=V_{G_j}\delta_{\boldsymbol{G}_{mn}-\boldsymbol{G}_{m'n'},\boldsymbol{G}_j},
\end{align}
whith amplitudes
\begin{align}
    V_{\boldsymbol{G}_{j}}=\begin{cases}
W_{j}, & U(r)=U(-r)\\
i(-1)^{j}W_{j}, & U(r)=-U(-r)
\end{cases}
\label{eq: Amplitudes}
\end{align}
dependent on the symmetry of the SL potential. Here, each $\boldsymbol{G}_j$ is a Fourier component defining the SL potential.\ In the first harmonic approximation they satisfy $|\boldsymbol{G}_j|=|\boldsymbol{G}_1|$ and $W_{j} = W$.\ To obtain the electronic structures, shown in Fig.~\ref{fig:Bandcharge} and Fig.~\ref{fig:BandTopo}, the resulting matrix is diagonalized by truncating the number of reciprocal lattice vectors until convergence.\ This methodology has been widely used to determine the electronic properties of graphene multilayers and transition metal dichalcogenides placed on top of a patterned dielectric SL~\cite{ghorashi2023multilayer,zeng2024gate,krix2023patterned,sun2023signature,ghorashi2023topological,yang2022chiral,shi2019gate,tan2024designing,zhan2024designing,Shi2025Fractional}. 

\section{Quantum Geometry} \label{topology}
In Eq.~\eqref{eq: Amplitudes} we define the amplitudes of the Fourier components depending on the symmetry of the SL potential.\ In the triangular and square cases studied before, the SL potential is inversion symmetric as $U(r)=U(-r)$. However, if we set  $\phi= \pm\pi/2$, Eq.~\eqref{eq: Ugeneral} takes the form
\begin{equation}
\begin{aligned}
U(\mathbf{r'}) &= 2 W \left[ \cos \left(\boldsymbol{G}_1 \cdot \mathbf{r'}\right) + \cos \left(\boldsymbol{G}_2 \cdot \mathbf{r'}\right) \right. \\  
&\left. + \cos\left(\boldsymbol{G}_3 \cdot \mathbf{r'}\pm\frac{3\pi}{2} \right) \right],
\end{aligned}
\label{eq: Ugeneralodd}
\end{equation}
which can be expressed as a sum of sine functions by returning to the original coordinates
\begin{equation}
     U(\mathbf{r})  =2 W \sum_{j=1}^{3} \sin \left(\mathbf{G}_j \cdot \mathbf{r}\right).
\label{eq: SinPote}     
\end{equation}
The above equation can be written as
\begin{equation}    U(\boldsymbol{r})=W\sum_{j=1}^{6}i(-1)^{j}e^{i\boldsymbol{G}_{j}\cdot\boldsymbol{r}},
\end{equation}
where $j$ runs over the six vectors defining the first harmonic functions. The above representation justifies the definition in Eq.~\eqref{eq: Amplitudes}.\ However, even though Eq.~\eqref{eq: Ugeneralodd} and Eq.~\eqref{eq: SinPote} describe the same potential, their different forms highlight the role of individual harmonics. Notably, this antisymmetric potential, shown in Fig.~\ref{fig:BandTopo}f), breaks inversion symmetry.\ This explains why the lower energy band (red line) in Fig.~\ref{fig:BandTopo} remains separated from the rest of bands regardless of the sign of $W$. Such potentials are known to induce non-trivial band topology in graphene heterostructures~\cite{Wallbank2013Generic,SanJose2014Electronic,SanJose2014Spontaneous}.

In SL with square, triangular or hexagonal geometries, a symmetric potential always results in a zero Chern number, with bands exhibiting zero Berry phase.\ However, when the potential is antisymmetric, although the total Chern number of the isolated bands remains zero, they develop a nonzero local Berry phase. This is summarized in Fig.~\ref{fig:BandTopo}, which shows the electronic structure and Berry phase, $\gamma_{\boldsymbol{k}}$, for the red- and green-colored bands.\ The Berry phase is calculated following the standard procedure detailed in Ref.~\cite{Xiao2010Berry} (see Appendix~\ref{Sec:BerryCurv}).\ For the red-colored bands, it is positive at the $K$ points and negative at the $K'$ points (see inset in Fig.~\ref{fig:Bandcharge}a for symmetry point definitions), closely resembling the band topology of gapped monolayer~\cite{Haldane1988Model} or bilayer graphene~\cite{Novoselov2006Unconventional}.  

Interestingly, we found that the non trivial quantum geometry originates from a specific pattern in the triangular superlattice, which is not explicitly visible in Eq.~\eqref{eq: SinPote} but becomes clear in Eq.~\eqref{eq: Ugeneralodd}.\ While two harmonics generate a periodic SL, the third can either preserve or break inversion symmetry.\ This reveals that the parity of the SL potential can be controlled by dephasing a single harmonic component, which can be achieved, for example, by encapsulating the 2DEG with two patterned gates or through remote proximity effects~\cite{Gu2024Remote, Zhang2024Engineering, He2024Dynamically}.\ Notably, this approach is similar to techniques used in optical lattices~\cite{Zheng2024Dynamical} but applied here to patterned systems. 

Moreover, the specificity of the pattern that gives rise to a non trivial Berry phase implies that the quantum dot limit is incomplete unless it is complemented by the lattice geometry. The subtle reason is that even an exponentially small overlap between the eigenfunctions of the quantum dots can carry information about relative phases—much like in the original Thouless picture of adiabatic topological phases~\cite{Thouless_1983}.\ That is, a very weak, low-frequency external driving of the parameters does not induce transitions to high-energy states.\ However, even a small projection onto a high-energy state can produce a measurable phase difference when a closed cycle is performed.\ This effect can be clearly seen in the potential studied in Appendix~\ref{Sec:AppLargeW} where the sine potential is gradually mixed with the cosine potential.\ This opens a gap in the Dirac cone seen in Fig.~\ref{fig:Bandcharge}a), Fig.~\ref{fig:Bandcharge}b) and Fig.~\ref{fig:Bandcharge}c), while the bands remain with nontrivial Berry phase in the limit of large $W$.

Finally, we note that a nonzero Berry phase also implies the presence of an orbital magnetic moment.\ Although the total Chern number of the low energy bands is zero, the Berry curvature has opposite signs at the $K$ and $K'$ points, closely resembling the case of gapped graphene~\cite{Zhang2005Experimental}.\ In the presence of a magnetic field, the Berry curvature couples to it, leading to corrections in both the anomalous velocity and the orbital magnetization~\cite{Fuchs2010Topological}.

\begin{figure}[ht]
\centering
\includegraphics[scale=0.52]{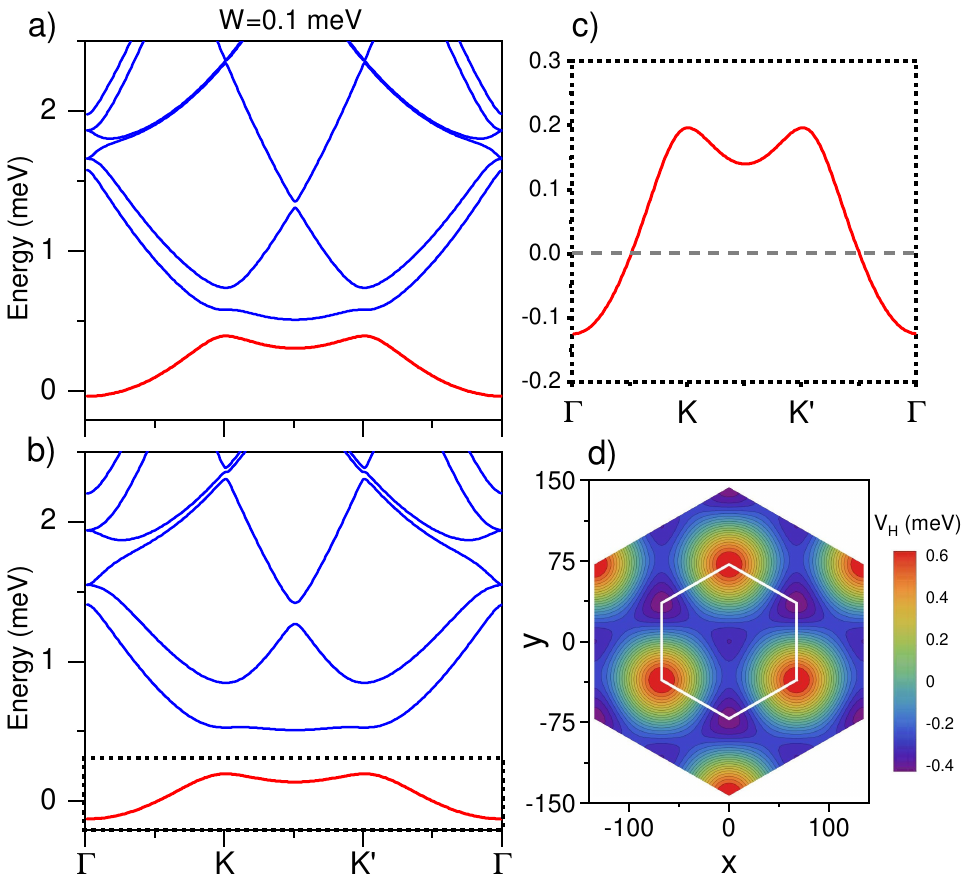}
\caption{Electronic structure of GaAs subjected to an antisymmetric scalar SL potential. In a) we show the bands without screening.\ In b) we show the self-consistent Hartree bands.\ Panel in c) is an enlarged region of the low energy band where the position of the Fermi energy is indicated by a gray line.\ By integrating the charge density from the bottom of the band to the Fermi energy we obtain the Hartree potential shown in panel d).}
\label{fig:Hartree}
\end{figure}

\section{Electrostatic Interactions} \label{electrostatic}
As shown in Fig.~\ref{fig:Bandcharge}, increasing the amplitude $W$ flattens the bands.\ The magnitude of this amplitude is determined by the gate voltages in the experimental setup~\cite{wang2024Tuning}.\ In the low-energy bands, as the band dispersion decreases, the wavefunctions at different momenta become progressively more alike, leading to a self-screening effect~\cite{Tkachenko2014Effects}.\ As the bands are filled, their contribution to the charge density can be expressed in terms of a parameter $\rho_H$, which captures the symmetric and antisymmetric components of the charge density.\ The value of $\rho_H$ depends on the wavefunction distribution in momentum space and can be computed self-consistently using a mean-field Hartree approximation.\ Following Refs.~\cite{Guinea2018Electrostatic,Cea2019Electronic}, this parameter is given by:
\begin{equation}
\rho_\text{H}(\textbf{G})=2 \int \frac{d^{2}\boldsymbol{k}}{A_\text{sBZ}}\sum_{\textbf{G}^{\prime},l}\psi_{k,l}^{\dag}(\textbf{G}^{\prime})\psi_{k,l}(\textbf{G}+\textbf{G}^{\prime}),
\label{eq: rho components} 
\end{equation}
where, ${A_\text{sBZ}}$ is the area of the SL Brillouin zone and $l$ is a band index. $\psi_{k,l}$ are the eigenvectors resulting from the self-consistent diagonalization of the Hamiltonian in Eq.~\eqref{eq: hamprincipal}.\ Because $\rho_H$ is in general a complex number, the Hartree potential in real space is given by~\cite{Guinea2018Electrostatic}
\begin{equation}
V_\text{H}(r)=2\sum_{mn}V_{0}\left(\textbf{G}_{mn}\right)\left|\rho_\text{H}(\textbf{G}_{mn})\right|\cos\left(\phi_{mn}+\textbf{G}_{mn}\cdot r\right),
\label{eq: FourierHartree}
\end{equation}
where  $V_{0}(\boldsymbol G)=v_C(\textbf{G})/A_{c}$, with $v_C(\textbf{G})=2 \pi e^2/(\epsilon |\textbf{G}|)$ is the Fourier transform of the Coulomb potential evaluated at $\textbf{G}$, $\bar{\epsilon}=10$ the dielectric constant, $A_{c}$ is the area of the SL unit cell.\ In this context, the phase  $\phi_{mn}= \arg\left[\rho_\text{H}(\textbf{G}_{mn})\right]$. The above equation implicitly depends on the filling fraction $\nu$, with $\nu=2$ for a full filled band. In addition, the variation of the charge density is such  that $\rho_H(G,\nu)=\rho_0+\delta\rho_H(G,\nu)$, we assume that $\rho_0=0$ at the bottom of the first band. 

\subsection{Symmetric Potential}

In the low energy regions the charge distribution is proportional to the SL potential but with a negative sign, as described in Sec.~\ref{nature}.\ This is illustrated in Fig.~\ref{fig:Bandcharge}d)-h) and Fig.~\ref{fig:Bandcharge}i) for the symmetric triangular and square SL, respectively.\ Since the Fourier components of the Hartree potential in Eq.~\eqref{eq: rho components} are determined by integrating the charge distribution from charge neutrality to a given filling, the Hartree potential is also proportional to the SL potential.\ However, the Hartree screening opposes the SL strength, resulting in an effective potential $V_{eff} = U + V_H$.\ This screening effect weakens the SL potential because, as $W$ increases, more charge accumulates in the attractive regions, generating a repulsive potential in those areas such that $U$ and $V_{H}$ have opposite signs.\ For example, by considering the triangular SL in Fig.~\ref{fig:Bandcharge}c) where $W=3.0$ meV.\ A Hartree correction with a filling fraction of $\nu=1.1$ results in $\rho_G =-0.35$ and $V_H =-2.0$ meV.\ The effective potential is then $V_{eff}=1.0$ meV which corresponds to the bands in Fig.~\ref{fig:Bandcharge}b).\ It is important to mention that for the considered SL potentials only the vectors with $|G_{mn}|=|G_1|$ are required in Eq.~\eqref{eq: FourierHartree}, in addition, the Hartree screening generated by the low energy bands is a real number and therefore $\phi_{mn}=0$.\ As the filling increases to the high energy bands the wavefunctions become too complex and additional Fourier components may be required~\cite{Wang2024Dispersion,wang2024Tuning}.

\subsection{Antisymmetric Potential}
We now analyze the case of an antisymmetric scalar potential.\ In this scenario, the potential in Eq.~\eqref{eq: FourierHartree} becomes complex due to the nonzero phases $\phi_{mn}$. Similar to graphene/hBN systems~\cite{Pantaleon2021Narrow}, the Hartree potential can be inferred from the charge distribution, which reflects the total charge density of a band at a given filling. As shown in Fig.~\ref{fig:BandTopo}g), the charge density at $\Gamma$ exhibits three maxima and three minima at the unit cell corners, with a small central distribution.\ A similar pattern appears at other high-symmetry points.\ Since this charge distribution lacks the symmetry of a purely triangular, Fig.~\ref{fig:Bandcharge}d), or honeycomb potential, Fig.~\ref{fig:Bandcharge}h), the Hartree potential, with a nonzero phase $\phi_{mn}$, acquires both symmetric and antisymmetric components.

Figure~\ref{fig:Hartree} compares the band structures without Hartree screening, Fig.~\ref{fig:Hartree}a), and with Hartree screening, Fig.~\ref{fig:Hartree}b). Figure~\ref{fig:Hartree}c) is an enlarged region of the Hartree bands where the position of the Fermi energy is indicated.\ As anticipated, the Hartree potential is complex, containing both even and odd components, shown in Fig.~\ref{fig:Hartree}d) where contour lines have been added for better visualization.\ This is unexpected, as one would assume the charge distribution in the bands to follow the purely odd symmetry of the SL potential as in the previous cases.\ We attribute this effect to subtle charge density variations across the sBZ, which introduce both even and odd components.\ Interestingly, these mixed odd/even potentials closely resemble those induced by an hBN substrates in graphene multilayers~\cite{Wallbank2013Generic,SanJose2014Electronic,SanJose2014Spontaneous}, but here they emerge from the interplay between antisymmetric SL potentials and electrostatic interactions.    

\section{Conclusions}
\label{Sec:Conclusions}

In this work, we analyzed the effects of periodic SL potentials and electrostatic interactions on 2DEGs. By transforming the Schr\"{o}dinger equation into ordinary second-order Mathieu equations, we derived exact analytical solutions describing the emergence of narrow bands and pseudo-Landau levels in 2DEGs under triangular and square SL potentials.\ Additionally, we provided a comprehensive phase diagram for the electronic structure as a function of the SL strength and complemented our results with numerical analysis using standard Bloch wave techniques.  

In the case of the triangular SL, we demonstrated that a non-trivial Berry curvature can emerge by breaking inversion symmetry through the relative phase between the harmonics of the potential.\ This implies that the quantum dot limit has important nuances and subtleties as needs to be complemented by the lattice geometry due to the small overlap between eigenfunctions that carry information about relative phases.\ Moreover, such mechanism enables topological control of electronic states without requiring external magnetic fields or spin-orbit coupling, broadening the possibilities for engineering topological phases in 2DEGs.  

By introducing a self-consistent Hartree potential, we uncovered a competition between the SL potential and screening effects, leading to a reduction of the effective potential in the symmetric case and the emergence of a mixed odd/even effective SL potential in the antisymmetric case.\ This interplay provides a mechanism to control and fine-tune electronic band structures through external periodic potentials. These findings enhance our understanding of SL-modulated systems and pave the way for engineering novel electronic phases in 2DEGs.  

\section*{Acknowledgments}
We thank Francisco Guinea for discussions and
Julian P. Ingham from Columbia University for providing useful references and insights on the subject.\ This work was supported by CONAHCyT project 1564464 and UNAM DGAPA project IN101924. IMDEA Nanociencia acknowledges support from the ‘Severo Ochoa’ Programme for Centres of Excellence in R\&D (CEX2020-001039-S/AEI/10.13039/501100011033).\ P.A.P acknowledges support from NOVMOMAT, project PID2022-142162NB-I00 funded by MICIU/AEI/10.13039/501100011033 and by FEDER, UE as well as financial support through the (MAD2D-CM)-MRR MATERIALES AVANZADOS-IMDEA-NC.\ Z.Z acknowledges support from the European Union's Horizon 2020 research and innovation programme under the Marie-Sklodowska Curie grant agreement No 101034431.

\appendix

\section{Ground state}\label{Sec:AppExpnasion}

In this Appendix we show that the ground state is proportional to the potential whenever $q_j<<1$. Let's consider Schr\"{o}dinger's equation with the Hamiltonian given by Eq.~\eqref{eq: total hamiltonian}, and for simplicity, we discuss only the case $|\mathbf{G}_j|=G$. It can be written as,
\begin{equation}
\label{eq:dimensiolessSho}
   -\frac{\nabla^2\psi(\mathbf{r})}{|G|^{2}} +\alpha w(\mathbf{r}) \psi(\mathbf{r})=\epsilon \psi(\mathbf{r}) 
\end{equation}
where we defined the adimensional potential,
\begin{equation}
    w(\mathbf{r})=U(\mathbf{r})/W,
\end{equation}
the effective strength interaction ratio,
\begin{equation}
    \alpha=\frac{2m^*W}{\hbar^{2}|G|^{2}},
\end{equation}
and the adimensional energy,
\begin{equation}
    \epsilon= \alpha=\frac{2m^*E}{\hbar^{2}|G|^{2}}.
\end{equation}
Since the SL potential $U(\textbf{r})$ is periodic in the lattice, then Bloch's theorem tells us that the solution of Schr\"{o}dinger's equation is given by $\psi_{\textbf{k}}(\textbf{r}) = e^{i \textbf{k} \cdot \textbf{r} } u_{\mathbf{k}}(\textbf{r})$, where $\textbf{k}$ is the crystal momentum vector and $u_{\mathbf{k}}(\textbf{r})$ is a periodic function in the lattice. Now we take a perturbative expansion in the adimensional parameter $\alpha$ over the wave function $u_{\textbf{k}}(\textbf{r})$ and the adimensional energy $\epsilon_{\mathbf{k}}$, this is,
\begin{align}
    u_{\textbf{k}}(\mathbf{r}) &= \sum_{n=0}  \alpha^{n} u_{\textbf{k}}^{(n)}(\mathbf{r}), \\
    \epsilon_{\textbf{k}} &= \sum_{n=0} \alpha^{n}\epsilon_{\textbf{k}}^{(n)},
\end{align}
where the index $n$ represent the order in $\alpha$. Since $ u_{\textbf{k}}(\mathbf{r})$ is periodic in the lattice, the functions $u_{\textbf{k}}^{(n)}(\textbf{r})$ are also periodic. Therefore, as they inherit the lattice periodicity we can Fourier expand them over the reciprocal lattice as
\begin{equation}
    u_{\textbf{k}}^{(n)} (\textbf{r}) = \sum_{\textbf{G} \in \Lambda.^*} e^{i \textbf{G} \cdot \textbf{r}}S_{\mathbf{k}}^{n}(\textbf{G}),
    \label{eq:basis}
\end{equation}
where $S_{\mathbf{k}}^{n}(\textbf{G})$ are the coefficients to be determined from the Bloch equation. In this particular case, we are interested only in the ground state which is obtained for the $\Gamma$ point, i.e., at $\mathbf{k}=(0,0)$. Also, Eq.~\eqref{eq:basis} suggests that the wave function has a similar structure to $U(\textbf{r})$. Thus it is natural to propose up to first order in $\alpha$ a wavefunction of the type,
\begin{equation}
\psi_{\mathbf{\Gamma}}(\textbf{r}) \approx 1+\alpha[1+\beta w(\textbf{r})],
\end{equation}
where $\beta$ is a constant to be fixed.\ Using Eq.~\eqref{eq:dimensiolessSho}, the zero order equation in $\alpha$ gives $\epsilon_{\mathbf{\Gamma}}^{(0)}=0$.
By collecting the first order terms in $\alpha$,
\begin{equation}
    (\beta+1)w(\mathbf{r})= \epsilon_{\mathbf{\Gamma}}^{(1)},
\end{equation}

leading to $\epsilon_{\mathbf{\Gamma}}^{(1)}=0$ and $\beta=-1$.
Finally, 
\begin{equation}
\psi_{\Gamma}(\textbf{r}) \approx C\left(1+\frac{2m^*}{\hbar^2|G|^2}(W-U(\textbf{r})) \right),
\end{equation} where $C$ is set by normalization over the Brillouin zone. On the other hand, the limit $q_j>>1$ corresponds to $\alpha \rightarrow \infty$. In that case we can use a boundary layer theory approach, i.e., if the Laplacian term occurring in Eq.~\eqref{eq:dimensiolessSho} is neglected, this leads to $\psi_{\Gamma}(\textbf{r})=0$ in most regions of the unitary cell. However, whenever
\begin{equation}
    -\frac{\nabla^2\psi(\mathbf{r})}{|G|^{2}} \sim \alpha,
\end{equation}
the whole equation needs to be taken into account. As a result, states have amplitude mainly around the minima of $U(\textbf{r})$, leading to the pseudo Landau level approach discussed before.

\begin{figure}
\centering
\includegraphics[scale=0.55]{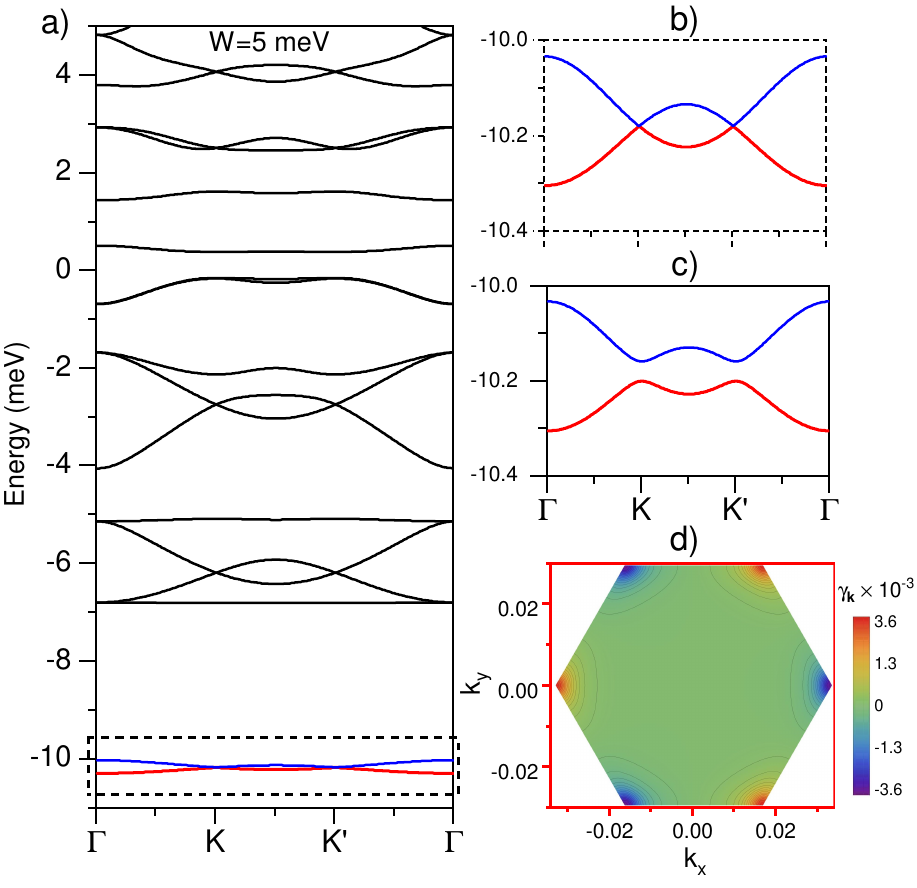}
\caption{Electronic structure of GaAs subjected to a SL potential with $W_1 = 5$ meV. In panel a), the band structure is shown. The inset in panel b) displays the low-energy bands for $W_2 = 0$, while panel c) shows the bands for $W_2 = 5\times10^{-3}$ meV, where a gap opens. Panel d) displays the local Berry phase acquired by the gapped low-energy, red colored band, of panel in c).}
\label{fig:LargeW}
\end{figure}

\section{Berry Curvature and Berry Phase}\label{Sec:BerryCurv}
In two-dimensional lattice systems, the Berry curvature of a band labeled by $l$ is given by~\cite{Berry84Quantal}
\begin{equation}
\label{eq:Omega}
\Omega_{\bm{k},l} = -2 \sum_{l' \neq l} \mathrm{Im} \left[ 
\frac{
\braket{\psi_{\bm{k},l}|\partial_{k_x} H_{\bm{k}}|\psi_{\bm{k},l'}} 
\braket{\psi_{\bm{k},l'}|\partial_{k_y} H_{\bm{k}}|\psi_{\bm{k},l}}
}{
(\epsilon_{\bm{k},l} - \epsilon_{\bm{k},l'})^2
} \right],
\end{equation}
where $\psi_{\bm{k},l}$ and $\epsilon_{\bm{k},l}$ are the eigenvectors and eigenvalues of the Hamiltonian $H_{\bm{k}}$. In our system, this quantity becomes non-zero when the SL potential is antisymmetric. The corresponding Chern number is defined as
\begin{equation}
\mathcal{C}_l = \frac{1}{2\pi} \int_{sBZ} \Omega_{\bm{k},l} \, d\bm{A},
\end{equation}
where sBZ denotes the SL Brillouin zone, and $d\bm{A}$ is an infinitesimal area element. The integral of the Berry curvature over the Brillouin zone gives the Berry phase $\gamma_l$, such that $2\pi \mathcal{C}_l = \gamma_l$. In the main text, we plot the local Berry phase, normalized by $2\pi$, given by
\begin{equation}
\gamma_{\bm{k},l} = \frac{1}{2\pi} \Omega_{\bm{k},l} \cdot d\bm{A}.
\end{equation}

\section{Quantum Geometry for Large W}\label{Sec:AppLargeW}
In the limit of large $W$, the electronic structure exhibits a pseudo-Landau level behavior. The low-energy bands become nearly flat, with their dispersion decreasing as $W$ increases. As discussed in the main text, even at large $W$ there remains an overlap between the narrow bands. To exemplify this effect, we consider a SL potential of the form
  \begin{equation}
       U(\mathbf{r}) = 2W_1 \sum_{j=1}^{3} \cos \left(\mathbf{G}_j \cdot \mathbf{r}\right)+2W_2\sum_{j=1}^{3} \sin \left(\mathbf{G}_j \cdot \mathbf{r}\right)\,
  \end{equation}
where $W_1$ and $W_2$ are the amplitudes of the even/odd components of the SL potential.\ For $W_1 > 0$, the SL potential acquires a honeycomb geometry (see Sec.~\ref{Sec:Hexa}), leading to two graphene-like low-energy narrow bands~\cite{wang2024Tuning}. As shown in Fig.~\ref{fig:LargeW}a) and in the inset of Fig.~\ref{fig:LargeW}b), these bands are degenerate at the $K$ and $K'$ points and become increasingly narrow for large $W_1$. Introducing the odd SL potential breaks inversion symmetry and opens a gap, as illustrated in Fig.~\ref{fig:LargeW}c) for $W_2 = 5\times10^{-3}$ meV.\ A  non-trivial local Berry phase emerges, as shown in Fig.~\ref{fig:LargeW}d).\ Remarkably, this topological feature persists even in the large $W_1$ limit, indicating that the wavefunctions retain information from nearby electronic bands.


%

\end{document}